\documentclass[twocolumn,showpacs,preprintnumbers,amsmath,amssymb,floatfix]{revtex4}

\usepackage{epsfig,psfrag}
\usepackage{dcolumn}
\usepackage{bm}
\usepackage{graphicx}
\usepackage{color}

\begin{document}

\title{Superconducting transport through a vibrating molecule}
\author{A. Zazunov,$^{1,2}$ R. Egger,$^{1,3}$ C. Mora,$^{1,3}$
and T. Martin$^{1,4}$} 
\affiliation{
${}^1$~Centre de Physique Th\'eorique, Case 907 Luminy, 
F-13288 Marseille cedex 9, France \\
${}^{2}$~LPMMC CNRS, 25 av. des Martyrs, F-38042 Grenoble, France \\
${}^3$~Institut f\"ur Theoretische Physik, Heinrich-Heine-Universit\"at,
D-40225  D\"usseldorf, Germany \\
${}^4$~Universit\'e de la M\'edit\'erann\'ee, F-13288 
Marseille cedex 9, France}

\date{\today}

\begin{abstract}
Nonequilibrium electronic transport through a molecular level
weakly coupled to a single coherent phonon/vibration mode has been
studied for superconducting leads.  The Keldysh Green function 
formalism is used to compute the current for the entire bias voltage range.
In the subgap regime, Multiple Andreev Reflection (MAR) processes
accompanied by phonon emission  cause rich structure near the
onset of MAR channels, including an even-odd parity
effect that can be interpreted in terms of an inelastic
MAR ladder picture. Thereby we establish a connection between 
the Keldysh formalism and the Landauer scattering approach 
for inelastic MAR.
\end{abstract}

\maketitle

\section{Introduction}

One of the primary goals in the field of molecular electronics is to 
understand quantum transport through individual nanoscale objects,
such as molecules, short carbon nanotubes, or DNA; 
for reviews, see Refs.~\cite{nitzan,book,tsukada}.
An important difference to conventional mesoscopic transport through quantum dots or 
granular islands arises because molecules can have intrinsic
vibrational degrees of freedom ('phonons') that may give rise to
Franck-Condon sidebands or phonon blockade in electronic transport.
Such features have been studied theoretically in many recent papers
\cite{schoeller,flensberg,oppen,mitra,cornaglia}.  
Molecular electronics is particularly exciting because 
it is in principle possible to contact molecules by leads of 
different nature.  Here we discuss how nonequilibrium transport
is affected by a coherent phonon mode coupled to the molecular charge
for the case of ($s$-wave BCS) {\sl superconducting leads}.  
Molecules connected to superconductors promise 
a rich terrain of exploration that allows for new 
spectroscopic tools (probing molecular properties), potentially
useful applications, and interesting fundamental physics.  
First experimental results have appeared for 
carbon nanotubes \cite{buitelaar,helene,reulet,leo,poul}
 and metallofullerenes \cite{kasumov}.  

So far transport through molecules has been theoretically
studied only for normal leads, either using rate equations 
or (in the quantum-coherent regime) perturbation theory in 
the electron-phonon coupling $\lambda$ \cite{mitra}.  
For superconducting leads and large transmission 
through the molecule, subgap transport is ruled by
MAR processes \cite{KBT,arnold}.  These have been extensively 
studied for point contacts \cite{bratus,averin,cuevas} 
and for junctions containing a resonant level \cite{alfredo,johansson}. 
In this paper, we provide a theoretical framework 
to include vibrations into superconducting transport
through a resonant molecular level. 
We focus on the most interesting quantum-coherent 
low temperature ($T$) limit with high (bare) transmission, where
Coulomb charging effects are largely wiped out. Therefore
the Coulomb interaction on the molecule will be neglected here.
We compute the d.c.~current basically for
the full bias ($V$) range within a 
Keldysh Green function scheme valid for small
$\lambda$ but arbitrary phonon frequency $\omega_0$.  
This approximation is current-conserving for identical molecule-lead
couplings $\Gamma_L=\Gamma_R=\Gamma$ and
superconductor gaps $\Delta_L=\Delta_R=\Delta$.

Our main results are as follows.
(i) For $eV\gg \Delta$, the coupling $\lambda$ tends to {\sl enhance}
the excess current $I_{exc}$, which is defined as the difference in
current for $V\to \infty$ when changing normal into superconducting 
leads.  (ii) In the subgap regime $eV< 2\Delta$,
the phonon gives rise to a quite complicated structure in the $I$-$V$
characteristics around each MAR onset at $eV=2\Delta/n$ (integer $n$),
with a pronounced {\sl even-odd parity} dependence.
These results can be qualitatively understood within a MAR
ladder picture.  Such a picture has previously been
used for junctions with a resonant level 
\cite{johansson} and is here extended to include 
phonon-induced transitions (inelastic MAR).  
Rich features in the $I$-$V$ curve 
appear already for $eV\ll\hbar\omega_0$, in contrast to 
normal leads where phonon signatures (e.g.~sidebands)
emerge only at $eV\ge \hbar\omega_0$ \cite{flensberg,mitra}.
(iii)  For $V=0$, we give analytical results for the
Josephson current in the adiabatic limit, 
$\hbar \omega_0 \ll \Delta \ll \Gamma$.
We find a  reduction (but no destruction) of the critical current 
and a changed current-phase relation.  These findings are
in qualitative agreement with Ref.~\cite{novotny},
where the opposite limit $\Gamma\ll \Delta$ has been studied
by lowest-order perturbation theory in the lead-molecule hopping.

The outline of the paper is as follows. In Sec.~\ref{sec2}, we discuss
the model and the Keldysh Green function approach taken in this work.
In Sec.~\ref{sec3}, we present our results and provide a physical
interpretation in terms of a MAR ladder picture.  
How such a scattering-type approach can be connected to the Keldysh approach
is explained in detail in the Appendix.  Finally, some
conclusions are offered in Sec.~\ref{sec4}. 
We put $e=\hbar=1$ in intermediate steps.

\section{Model and Keldysh approach}
\label{sec2}

\subsection{Model}

We now wish to formulate and analyze a tractable model describing the relevant physics 
of a molecule sandwiched between superconducting leads. 
Writing the Hamiltonian as
\begin{equation} \label{e1}
H = \omega_0 b^\dagger b  + \sum_{\sigma}
(\epsilon_0+\lambda X)  d^\dagger_\sigma d^{}_\sigma + H_L + H_R + H_T,
\end{equation}
we consider one relevant molecular level associated with the fermion operator $d_\sigma$ for
spin $\sigma=\uparrow,\downarrow$ and located at the energy $\epsilon_0$.
In Eq.~(\ref{e1}) we take a linear coupling 
between the molecular charge and 
the phonon displacement $X=b+b^\dagger$, where the boson operator $b$ annihilates a phonon
excitation.  For a justification of this 
form and possible other couplings, see Ref.~\cite{flensberg}.
The leads are described by a pair of standard $s$-wave BCS Hamiltonians.
Using the Nambu vector $\Psi^T_{j,k}=(\psi_{j,k,\uparrow}, 
\psi^\dagger_{j,-k,\downarrow})$ for electrons in lead $j=L/R$, we have
\begin{equation} \label{h2}
H_{j=L/R} = \sum_k \Psi_{jk}^\dagger 
\left (\begin{array}{cc}  \xi_k  & \Delta \\ \Delta^\ast & -\xi_k
 \end{array} \right) \Psi_{jk} 
\end{equation}
with single-particle energy $\xi_k=k^2/2m-\epsilon_F$; the  $2\times 2$
 matrix acts in Nambu space.
In the following, standard Pauli matrices in Nambu space are used and denoted
by $\sigma_{x,y,z}$.  Using the Nambu vector $d=(d^{}_\uparrow, d^\dagger_\downarrow)^T$
and $\Gamma=\pi\nu_0 |t_0|^2$ for
(normal) lead density of states $\nu_0$, the lead-molecule coupling is
\begin{equation}\label{h3}
H_T = t_0 \sum_{k,j=L/R=\pm} \Psi^\dagger_{jk} \sigma_z e^{\pm i\sigma_z V t/2} d + {\rm h.c.},
\end{equation} 
where the voltage $V$ enters via the time-dependent phase.
As we are mostly interested in the $V\ne 0$ case,
for simplicity, we consider $\Delta>0$ to be real-valued.
(In the study of the Josephson current, the residual $V=0$ phase difference across the
molecule is of course taken into account.) 

Note that Eqs.~(\ref{h2}) and (\ref{h3}) assume symmetric
coupling and identical superconducting gaps. 
Our approximation scheme below will yield a current-conservating result only then.
In fact, since the calculation of MAR-dominated transport is already 
involved for $\lambda=0$, a nontrivial current-conserving 
self-consistent approach covering the large transmission limit 
seems out of question. Note that self-consistency is (usually)
sufficient  to ensure current conservation \cite{baym}.
  Below we instead proceed using a perturbation theory with the 
small expansion parameter
$\lambda/\Gamma$.  Under such an approach,  current
conservation is known to only hold for the electron-hole symmetric case
\cite{hershfield}, i.e., $\epsilon_0=0$, with symmetrically arranged 
leads ($\Delta_L=\Delta_R=\Delta, \Gamma_L=\Gamma_R=\Gamma$).
This case is taken in what follows.
We note in passing that this important issue (and also the tadpole diagram
in Fig.~\ref{fig1}) was overlooked in
Ref.~\cite{mitra}, where the same approximation as ours was 
implemented for normal leads but also used for asymmetric cases. 

\subsection{Keldysh approach}

\begin{figure}
\scalebox{0.18}{\includegraphics{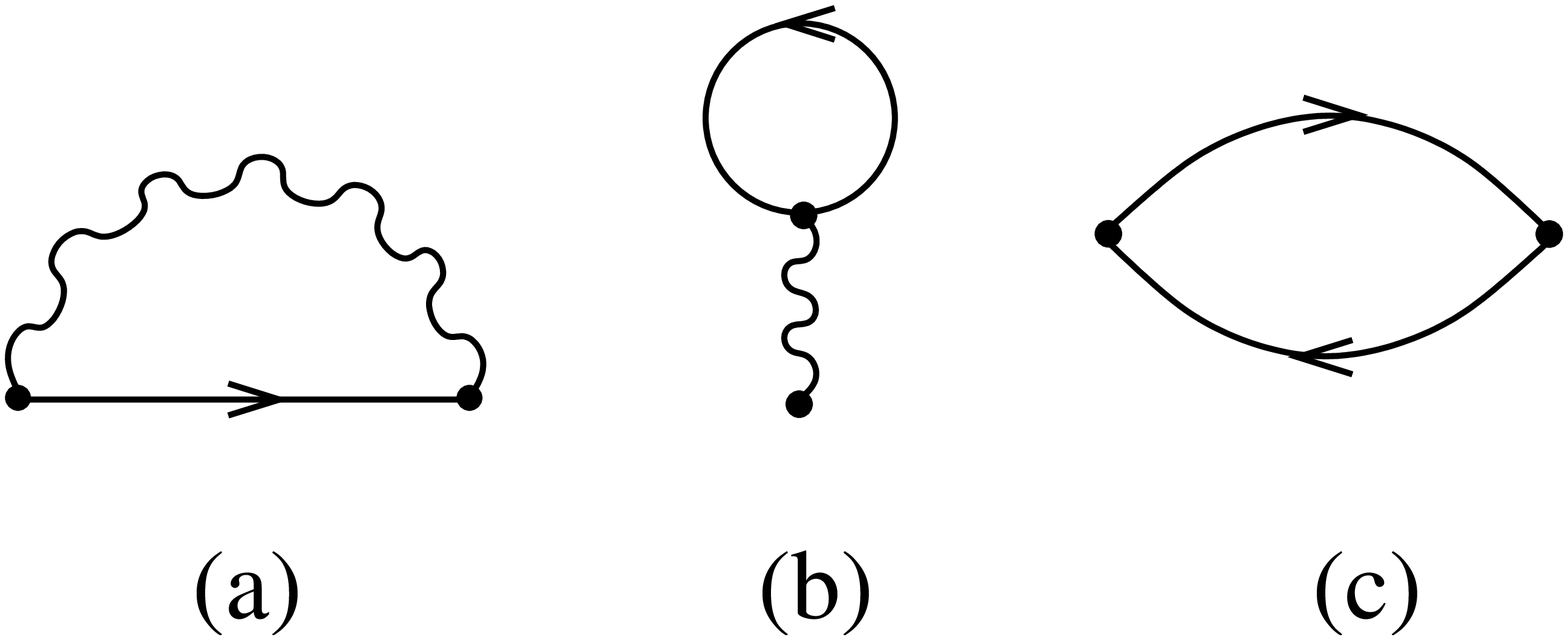}} 
\caption{ \label{fig1}
Self energy due to the presence of the phonon mode:
(a) 'Fock' and (b) 'tadpole' diagram.  The polarization bubble (c) 
leads to the dressed phonon propagator $\check{\cal D}$ (wiggly lines).
Arrowed lines denote $\check{\cal G}_0$.  }
\end{figure}

To compute the current-voltage characteristics,
we now employ the Keldysh Green function technique.  This
method has become a standard approach by now; 
for a review, see, e.g., Ref.~\cite{kamenev}.
The  Keldysh Green function for the $d$ fermion is defined as  
\begin{equation}\label{gd}
{\cal G}^{s s'}_{\alpha \alpha'}(t,t')=
-i\langle \hat T_C [ d_\alpha(t_s) d_{\alpha'}^\dagger (t'_{s'}) ]\rangle,
\end{equation}
where $\alpha,\alpha'$ ($s,s'=\pm$) are Nambu (Keldysh) indices, 
and $\hat T_C$ is the time-ordering operator along the familiar Keldysh 
contour $C$ \cite{kamenev}. Accordingly, $t_s$ denotes a time
taken on branch $s$ of the Keldysh contour.
Similarly, we define a phonon Keldysh Green function
${\cal D}^{s s'}(t,t')$ for the quantity $X= b+b^\dagger$
coupling to the fermion $d$ in Eq.~(\ref{e1}).
In what follows, we use the check notation ($\check{\cal G}$) in order to
schematically indicate the Keldysh structure. In the case of electron 
Green functions, this also includes the Nambu structure.
Denoting the respective functions for $\lambda=0$ 
by $\check{\cal G}_0$ and $\check{\cal D}_0$, and using the 
self-energy diagrams in Fig.~\ref{fig1},
the dressed Green functions used in our perturbative
approximation follow from the Dyson equation.
As we consider only the case $\epsilon_0=0$, it can be checked that
the 'tadpole' diagram does not contribute.

For the superconducting problem of interest here, it is
convenient to use the double Fourier representation 
\begin{equation}
\check{{\cal G}}(t,t') = \sum_{n, m = -\infty}^{+\infty} 
\int_F \frac{d \omega}{2 \pi} \, 
e^{-i \omega_n t + i \omega_m t'} \, \check{{\cal G}}_{nm}(\omega) ,
\end{equation}
and likewise for all other Green functions and self energies.
Here we use
\begin{equation}\label{wn}
\omega_n = \omega + n V
\end{equation}
for $\omega$ within the 'fundamental' domain $F$ defined as 
\begin{equation}\label{fun}
F\equiv [-V/2,V/2]. 
\end{equation} 
For fixed $\omega\in F$,
the Dyson equations then lead to the matrix equations
\begin{eqnarray} \label{g0}
\check{{\cal G}}_{0,nm}^{-1}(\omega) & = & \delta_{nm} 
\omega_n \check{\tau}_z - \Gamma \sum_{j=L/R} \check{\gamma}_{j,nm}(\omega) 
\\ \label{g1} \check{{\cal G}}^{-1} &=& \check{\cal{G}}_{0}^{-1} -
\check{\Sigma}, \quad \check{{\cal D}}^{-1}= \check{\cal{D}}_{0}^{-1} - \check{\Pi},
\end{eqnarray}
where the Pauli matrix $\check{\tau}_z$ acts in Keldysh space, and 
\begin{equation}
 \check{{\cal D}}^{-1}_{0,nm}(\omega) = \delta_{nm} \check{\tau}_z \frac{ 
 \omega_n^2 - \omega_0^2 }{2\omega_0}.
\end{equation}
The self energy due to tracing out the respective lead is given by the Nambu matrix
\begin{eqnarray} \label{selfen}
&& \check{\gamma}_{j=L/R=\pm,nm}(\omega) = \\ \nonumber
&& \left( \begin{array}{cc} 
\delta_{nm} \, \check{X}(\omega_n \mp V/2) & 
\delta_{m, n \mp 1} \, \check{Y}(\omega_n \mp V/2) \\ 
\delta_{m, n \pm 1} \, \check{Y}(\omega_n \pm  V/2) & 
\delta_{nm} \, \check{X}(\omega_n \pm V/2)
\end{array} \right) 
\end{eqnarray}
with Keldysh matrices $\check{Y}(\omega) = - \Delta\check{X}(\omega)/\omega$
and
\[
\check{X}(\omega) =   \left\{ \begin{array}{ll} 
- \frac{\omega}{\sqrt{\Delta^2 - \omega^2}} \check{\tau}_z ,
 & |\omega|<\Delta\\
 \frac{i|\omega|}{\sqrt{\omega^2-\Delta^2}} \left( \begin{array}{cc}
2 f_\omega - 1 & - 2 f_\omega \\ 2 f_{-\omega} & 2 f_\omega - 1
\end{array} \right) , & |\omega|>\Delta \end{array}\right. 
\]
where $f_\omega=1/(1+e^{\omega/k_B T})$ is the Fermi function.

Figure \ref{fig1} yields for
the polarization $\check{\Pi}$ and the self energy $\check{\Sigma}$
the following explicit expressions:
\begin{eqnarray}\nonumber
\check{\Pi}^{ss'}_{nm} (\omega) &=& -i \lambda^2 \ {\rm tr} \sum_{n' m'} 
\int_F \frac{ d\omega' }{2\pi} \left (\check{\tau}_K 
\check{\cal G}_{0;n'm'}(\omega') \check{\tau}_K \right)^{ss'}
\\ \label{s0} &\times& \check{\cal G}_{0; m'-m,n'-n}^{s's}(\omega'-\omega) ,
\\ \nonumber
\check{\Sigma}^{ss'}_{nm} (\omega) &=& i \lambda^2 \sum_{n' m'} 
\int_F \frac{d \omega'}{2 \pi}  {\cal D}^{ss'}_{n-n',m-m'}(\omega-
\omega') \\    &\times&             \label{s1}
\left( \check{\tau}_K
\check{\cal G}_{0; n' m'}(\omega')  \check{\tau}_K \right)^{s s'} ,
\end{eqnarray}
where 'tr' extends over Nambu space only and
 $\check{\tau}_K=\check{\tau}_z \sigma_z$.
Should the difference $\omega-\omega'$ 
appearing in Eqs.~(\ref{s0}) and (\ref{s1}) 
fall outside the fundamental
domain $F$, one has to fold it back to $F$.
This is implicitly understood in the above equations.

The steady-state d.c.~current through the left/right junction 
then follows in the form
\begin{equation} \label{dc_current}
I_{L/R} = \mp  2\Gamma\ {\rm Re} \sum_{nm} \int_F \frac{d \omega}{2 \pi} 
{\rm tr}\left[ \sigma_z\check{\gamma}_{L/R,nm}(\omega)
 \check{\cal G}_{mn}(\omega) \right]^{+-}.
\end{equation}
This relation constitutes a generalization of the Meir-Wingreen
formula \cite{mw} to the case of superconducting leads. 
We also define the phonon contribution 
\begin{equation}\label{iph}
\delta I_{ph}\equiv I(\lambda)-I(\lambda=0).
\end{equation}
Current conservation, $I_L=I_R\equiv I$,
can be explicitly verified as follows.
In the particle-hole symmetric case, current conservation requires that
\begin{equation}\label{rel}
\check {\cal G}^{ss'}_{nm}(\omega) =
 -\sigma_z  \check{\cal G}^{s's}_{mn}(-\omega) \sigma_z.
\end{equation}
This relation is indeed obeyed since the approximate $\check \Sigma$ in 
Eq.~(\ref{s1}) also fulfills Eq.~(\ref{rel}).

Using Eq.~(\ref{dc_current}), we evaluate the $I$-$V$ characteristics
for $\lambda=0.15 \Gamma$ (unless noted otherwise) and $k_B T/\Delta=0.01$.  
We truncate the summations such that $|\omega_n|<\omega_c = 20\Delta$;
further increase of the bandwidth $\omega_c$ did not change results. 
In practice, the domain $F$ in Eq.~(\ref{fun}) must be discretized.
Typically, we found $\delta\omega=0.008\Delta$ to be
sufficient for convergence. The matrix inversions
in Eqs.~(\ref{g0}) and (\ref{g1}) are then done
for each $\omega\in F$ separately, involving matrix dimensions
of the order $\omega_c/eV$.  For very small $eV/\Delta$, this becomes quite costly, and we 
limit ourselves to $eV/\Delta> 0.15$ in the following.

Fortunately, there are several nontrivial tests that we can use to check the scheme.
For $\lambda=0$, our approach quantitatively reproduces the results of 
Refs.~\cite{bratus,cuevas,johansson}.  For $\Delta=0$, we recover results of 
Ref.~\cite{mitra} when applicable (i.e., for $\epsilon_0=0$).  
As additional check, Green function sum rules \cite{kamenev} were verified,
such as 
\begin{equation}
{\rm tr}\left( \tau_K \check {\cal G}(t,t) \right)=0
\end{equation}
 at coinciding times.  While such relations
must hold for the exact Green function, it is reassuring to
verify that our approximation does not lead to violations of 
sum rules.

Besides the current, we have  also monitored the average
phonon number $N_{ph}=\langle b^\dagger b \rangle$ and the 
frequency-dependent phonon distribution function.
We find $N_{ph}\alt 1$ for $\lambda/\Gamma=0.15$, in accordance
with our assumption of weak electron-phonon coupling.
The phonon distribution function revealed renormalizations
of the peak position away from $\omega_0$ by a few percent.

To conclude this section, we note that it is possible to establish a close
connection between this Keldysh Green function approach and a Landauer scattering
approach incorporating inelastic transitions.  Such an 'inelastic MAR' picture
will in fact be essential in interpreting our numerical results in Sec.~\ref{sec3a}.
This connection is detailed in the Appendix.

\section{Results} \label{sec3}

\subsection{Subgap regime: Inelastic MAR}\label{sec3a}

\begin{figure}[t!]
\scalebox{0.28}{\includegraphics{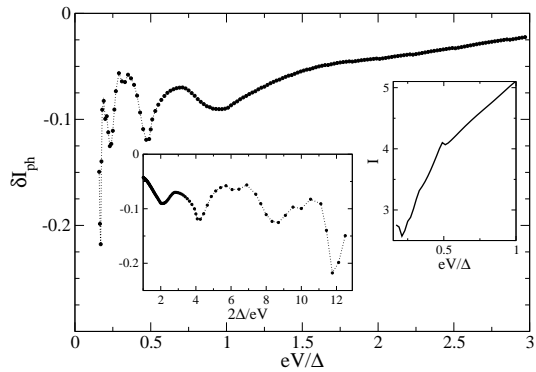}} 
\caption{ \label{fig2} Phonon difference current (\ref{iph}) 
for $\hbar\omega_0=0.2\Delta$ and $\Gamma=2\Delta$.  In all figures, 
currents are given in units of $e\Delta/(2\pi\hbar)$,
and dotted lines are guides to the eye only.
Left inset: Same as function of $2\Delta/eV$.
Right inset: Part of the total $I$-$V$ curve (note the scales). }
\end{figure}

\begin{figure}[t!]
\scalebox{0.25}{\includegraphics{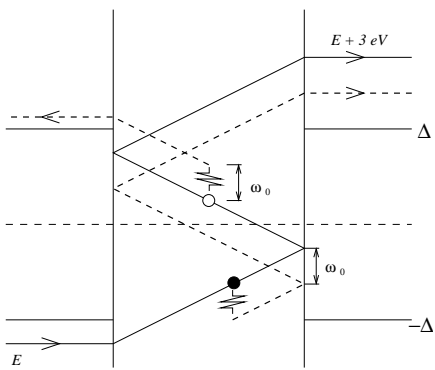}} 
\caption{ \label{fig3}
MAR ladder picture with phonon emission.  Here $eV$ is
slightly below $\Delta$: for an electron incoming from the left
side, we have one hole (open circle) 
and two electron (filled circle) segments.
Dashed lines indicate possible trajectories after single phonon
emission involving either hole or electron segments.
There is also a MAR path (not shown) for a hole
entering from the right side, with one electron and two hole segments.  }
\end{figure}

\begin{figure}[t!]
\scalebox{0.28}{\includegraphics{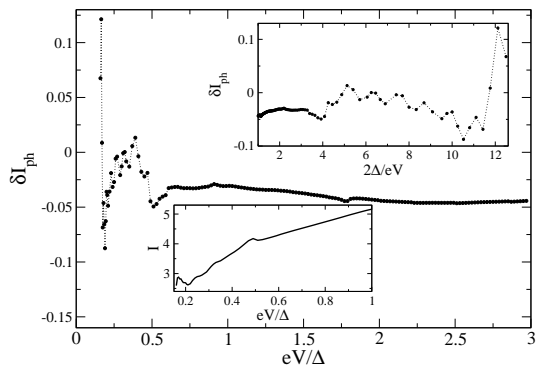}} 
\caption{ \label{fig4}
Same as Fig.~\ref{fig2} but for $\hbar\omega_0=1.8\Delta$. 
The upper inset gives $\delta I_{ph}$ as a function of $2\Delta/eV$,
the lower inset gives the low-voltage part of the total current.}
\end{figure}

Let us then turn to results for the $I$-$V$ curve. We start with
the subgap regime, where MAR provides the dominant transport mechanism. 
In particular, for $2\Delta/(n+1) < eV < 2\Delta/n$ ($n$ integer),
there is a total number $n$ of Andreev reflections for electrons or holes 
within the superconducting gap. 
The $I$-$V$ curve for $\Gamma=2\Delta$ and $\omega_0=0.2\Delta$ is
given in Fig.~\ref{fig2}, where $\delta I_{ph}$ is always negative. 
Note that in this fully transmitting limit, the $I$-$V$ curve for $\lambda=0$ 
is smooth and does not exhibit the MAR 'cusps' encountered at lower
transmission \cite{bratus}. Phonons now {\sl restore} 
such features near MAR onsets, with pronounced
{\sl even-odd 'parity' effects}: For even (odd) $n$, $\delta I_{ph}$ 
shows valleys (peaks) around $eV=2\Delta/n$. 
This is clearly seen in the left inset of Fig.~\ref{fig2} for
 $n$ up to 12 \cite{foot}. Note that the dip at $n=8$ is somewhat
shifted, presumably due to phonon renormalization effects.
However, the appearance of even-odd parity oscillations is quite
distinct and surprisingly regular given the complexity of this 
system.  

In order to achieve a physical understanding of this
even-odd effect, it is useful to
invoke a MAR ladder picture in energy space, including
inelastic transitions caused by phonon emission.
A schematic description of the MAR ladder picture in this limit
is given in Fig.~\ref{fig3}. Superconductor spectra 
are placed next to each other,
but because of the presence of the bias,
electrons emitted from left to right 
gain an energy $eV$ in this process. Holes that are reflected,
traveling from right to left, also gain this energy.

Usually the scattering approach is used to develop this picture
\cite{johansson}, but its straightforward implementation
 encounters conceptual difficulties in the presence of 
inelastic phonon transitions. Fortunately, a different route
resolving these difficulties can be formulated via the above Keldysh 
approach, whose formal justification is detailed in the Appendix.
Here electron and hole propagators in energy space
are coupled to each other through suitable Andreev reflection matching conditions
and the phonon self energy $\check{\Sigma}$.  
For a qualitative explanation of the even-odd effect
found in the full calculation, cp.~Fig.~\ref{fig2}, it is sufficient
to restrict the full MAR ladder picture to {\sl single} phonon
emission processes, where electrons (holes) lose (gain)
the energy $\hbar \omega_0$.  Since $N_{ph}\alt 1$, emission 
dominates over absorption and multi-phonon processes are rare.

In Fig.~\ref{fig3}, the two superconductors 
are positioned at the same chemical potential, but electrons
(from left to right) and holes (from right to left) `climb' the MAR ladder 
by gaining $eV$ for each Andreev reflection. 
The higher the total number of Andreev reflections in one cycle, 
the larger the total charge transmitted.
Since we consider the high transmission limit where
high-order MAR processes are not penalized,
the current is therefore expected to increase (decrease) if  phonon emission
is able to increase (decrease) the number of Andreev reflections in a MAR cycle.
For $eV$ slightly below  $2\Delta/n$ with {\sl even} $n$,  
we then argue as follows; for $n=2$, see
Fig.~\ref{fig3}.  For small energy transfer $\hbar \omega_0$, 
if a phonon is emitted during an electron segment,
MAR trajectories in energy space are not drastically modified
in the sense that the number of Andreev reflections stays unaffected. 
However, if a phonon transition occurs during a hole segment,
the MAR ladder is shifted upwards by $\hbar\omega_0$ and the
last hole on the MAR ladder can be scattered
into the continuum (left electrode in Fig.~\ref{fig3}) instead
of being Andreev reflected.  Consequently, one Andreev reflection is lost 
and hence the current is expected to decrease. 
This argument applies both to incoming electrons and holes, 
and explains why current valleys are observed for 
$eV\approx 2\Delta/n$ with even $n$ in Fig.~\ref{fig2}.
On the other hand, consider $eV$ slightly above $2\Delta/n$ with 
{\sl odd} $n$.  Reiterating the above analysis, now phonon emission during 
a hole segment tends not to affect the number of Andreev reflections.
If the phonon is emitted during an electron segment,
however, an additional Andreev reflection has to take place to 
complete the MAR cycle, leading to a 
current peak for $eV\approx 2\Delta/n$ with odd $n$.

Next consider  $\hbar\omega_0=1.8 \Delta$ but otherwise identical
parameters, see Fig.~\ref{fig4}, where $\delta I_{ph}$ 
can be positive and again shows oscillations near the MAR onsets, which are  
less pronounced for small $n=2\Delta/eV$.  Remarkably, even for small voltages, 
$eV\ll \hbar\omega_0$, a rather complicated subgap
structure is caused by the phonon.  At such low voltages,
this is only possible via MAR, for otherwise
electrons or holes do not have enough energy to emit
a phonon. The broad minimum corresponding to $n=2$
observed in Fig.~\ref{fig2} has now vanished.   
Let us invoke the MAR ladder picture to rationalize this effect.
For $eV\alt \Delta$, by emitting a {\sl high-energy} phonon ($\hbar\omega_0
> \Delta$), the last electron on the MAR ladder can now be 
scattered back inside the gap instead of heading to the 
continuum.  This increases the number of reflections and thus 
the current. As a phonon emitted during the hole
segment has the opposite effect (see above), the net outcome
of the higher phonon frequency is to suppress 
the valley at $eV\approx\Delta$ expected for small $\omega_0$.
Figure \ref{fig4} also shows a dip in the current at $eV\approx 1.8\Delta$,
representing a phonon backscattering feature at $eV=\hbar \omega_0$.
The scaling of this dip with $\omega_0$ was confirmed by additional 
calculations. One can also see a two-phonon feature
at $eV=\hbar\omega_0/2$ in Fig.~\ref{fig4}.

\begin{figure}[t!]
\scalebox{0.28}{\includegraphics{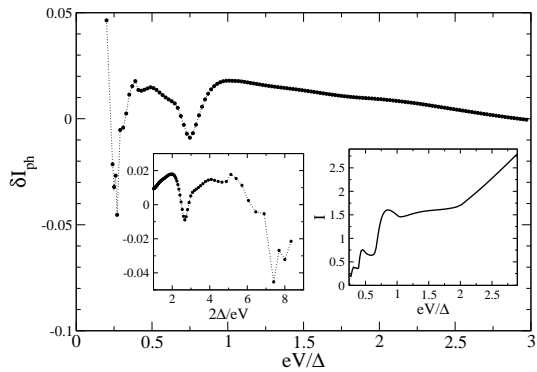}} 
\caption{ \label{fig5}
Same as Fig.~\ref{fig2} but for $\Gamma=0.5\Delta$. 
The left inset gives $\delta I_{ph}$ versus $2\Delta/eV$,
the right inset the total current.}
\end{figure}

Let us then briefly go back to $\hbar\omega_0=0.2\Delta$ (cp. Fig.~\ref{fig2}),
but now for $\Gamma=0.5\Delta$, see Fig.~\ref{fig5}. For small $\Gamma$, 
quasi-resonances appear \cite{johansson} and cause additional features,
e.g.~negative differential conductance portions in the $I$-$V$ curve.
The MAR ladder picture then has to include both 
quasi-resonances (cp. Ref.~\cite{johansson}) and phonon
transitions, which is possible but beyond the scope of this paper.
Fig.~\ref{fig5} shows that the even-odd parity effect requires
large $\Gamma$ to be observable.

\subsection{Excess and Josephson current}

Finally, we  briefly discuss the limits of very large and/or zero voltage.  
Starting with the first case (excess current), 
we have computed the difference $\delta I_{exc,ph}$ between the 
excess currents $I_{exc}$ with
and without the phonon. This calculation has been done  at $eV=10\Delta$, which
according to the discussion in Ref.~\cite{johansson} is certainly large enough 
for our parameters.  For the case of high transmission encountered here, 
we find that phonons generally {\sl enhance} the excess current.
To give a concrete example, for $\hbar\omega_0=0.8\Delta$, $\Gamma=2\Delta$
and $\lambda=0.5\Delta$, we find $\delta I_{exc,ph}/ I_{exc} \approx 0.07$.
A similar current enhancement at high transmission
was also found for environmental Coulomb blockade in 
superconducting junctions,
and has been explained as 'antiblockade' effect \cite{levy}.
As such, this effect of the phonon may not be too surprising.

Second, the equilibrium Josephson current has been calculated by
adopting our approach to the Matsubara representation. 
For arbitrary parameters, it is straightforward to
numerically compute the full current-phase relation.  However,
the current-phase relation in the adiabatic phonon regime defined by 
\begin{equation}
\hbar\omega_0 \ll \Delta \ll \Gamma
\end{equation} 
can even be calculated analytically, with the result
\begin{equation} \label{josephson_curr}
I(\phi) =  (e\Delta^2/2\hbar)\  T_0\sin (\phi) / E_0(\phi),  
\end{equation}
where 
\begin{equation}
E_0(\phi) = \Delta [1 - T_0\sin^2(\phi/2)]^{1/2}
\end{equation}
is an Andreev bound state energy \cite{bratus} 
in the junction with an effective transparency 
\begin{equation}
T_0=\frac{1}{1+\lambda^2/4\Gamma^2}.
\end{equation}
The $\phi$-dependent shift without any broadening of the Andreev
level caused by the coupling to a phonon mode is characteristic
for the coherent limit and decreases the critical current.
Very similar results were reported in Ref.~\cite{novotny}
in the opposite limit $\Gamma\ll \Delta$.

\section{Conclusions}\label{sec4}

In this work, we have for the first time theoretically explored 
nonequilibrium molecular transport with 
superconducting leads in the coherent phonon regime.
Phonons reveal a rich subgap structure 
even for voltages well below the phonon frequency, including
a surprising even-odd parity effect near the MAR onsets.
This effect can be largely understood within a physically
appealing inelastic MAR ladder picture, based on the assumption
that single phonon emission processes dominate.  For
stronger electron-phonon couplings, also multiple phonon 
processes and/or absorption becomes important, and the
practical usefulness of such a scattering approach is less clear,
although it can be formally derived as outlined in the Appendix.

Let us also offer  a brief outlook.
Besides the obvious interest to experimentally probe the
effects predicted here, the problems raised above
deserve further theoretical work.  In the case of strong
electron-phonon coupling, different approximations will
be necessary.  Moreover, other quantities such as 
shot noise or frequency-dependent noise \cite{noise} deserve 
attention.  Noise can yield information about the effective
charge involved in the transfer process, and how this charge
is affected by phonon transitions remains to be explored.
It would also be interesting to formalize the notion
of inelastic MAR spectroscopy, potentially allowing to infer
the electronic structure of molecules from the 
superconducting current-voltage characteristics.
Such approaches have already been very useful in nanoscale break junctions, 
and are of obvious interest in the molecular electronics context as well.
It is safe to conclude that superconducting molecular transport definitely
warrants further surprises in the future.

\acknowledgements

We thank K. Flensberg, A. Levy-Yeyati, and V. Shumeiko for useful discussions.
R.E.~and C.M.~are grateful to the CPT for hospitality.
R.E.~was supported by the CNRS, the ESF network INSTANS,
 and by the EU networks HYSWITCH and DIENOW.  
T.M. acknowledges support from an A.C. Nanoscience from CNRS.

\appendix
\section*{Inelastic MAR: Scattering approach}

In this appendix, we outline the derivation of the Landauer scattering approach to 
inelastic MAR incorporating phonon transitions.  Such an approach is formulated here using 
the equation-of-motion method, where transfer matrices are expressed in 
terms of Keldysh Green functions. 
This provides a formal justification for the intuitive MAR ladder picture
used in Sec.~\ref{sec3a}. 

Using the Hamiltonian (\ref{e1}), we can write down 
the equations of motion for the quasiclassical envelope  function
$\Psi_{j,k}^{(r)}$ describing right/left-moving quasiparticles ($r=\pm$)
in the left or right lead ($j=L/R=\pm$)
with momentum $rk_F+k$.  Here $|k| \ll k_F$ is assumed, and with the 
Fermi velocity $v_F$, we obtain
\begin{equation} \label{eom1}
(i \partial_t - rk v_F \sigma_z - \Delta \sigma_x )
\Psi_{j=\pm,k}^{(r)}(t) = t_0 \sigma_z e^{\pm i \sigma_z Vt/2} \  d(t) .
\end{equation}
Similarly, the equation of motion for the molecular fermion (Nambu spinor $d$) is
\begin{equation} \label{eom2}
(i \partial_t - \lambda X(t) \sigma_z )
d(t) = t_0 \sigma_z \sum_{j=\pm,r} e^{\mp i\sigma_z Vt/2}
 \Psi_j^{(r)}(t) ,
\end{equation}
where $\Psi_j^{(r)} \equiv \sum_k \Psi_{jk}^{(r)}$.
Here we consider the simplest possible situation, where phonon
renormalization effects like the polarization bubble in
Fig.~\ref{fig1} are completely disregarded.  In that 
case it is sufficient to simply take an equilibrium average
over the phonon subsystem, while otherwise
one should also take into account the equation of motion
for $X(t)$.  

 The solution of Eq.~(\ref{eom1}),  
 adapted to  $\Psi_{j}^{(r)}(t)$, can be written as the sum of a
 free (homogeneous) solution $\Psi_{j0}^{(r)}(t)$, describing an incoming quasiparticle 
in lead $j=L/R$, plus a scattered part due to the interaction with the
molecular level, 
\begin{eqnarray} \label{response}
\Psi_j^{(r)}(t) & = & \Psi_{j0}^{(r)}(t) + t_0
\int dt' \ g_j^{(r)} (t - t')  
\\ \nonumber && \times \sigma_z e^{\pm i\sigma_z Vt'/2} d(t') ,
\end{eqnarray}
which is expressed in terms of the retarded Green function $g_j^{(r)}$
for uncoupled electrodes whose Fourier transform is given by
\[
g_{L}^{(\pm)}(\omega) = g_R^{(\mp)}(\omega) =  
\frac{\pi\nu_0}{2i} \left ( \frac{\omega+\Delta\sigma_x}
{\sqrt{(\omega+i0)^2 -\Delta^2} } \mp \sigma_z \right) .
\]
When summed over $r$, this essentially yields
the retarded components of $\check{\gamma}_j$ in Sec.~\ref{sec2}. 

In what follows, we use $s = \{ 1, 2, 3, 4 \}$ in order to label scattering processes
corresponding to electron- or hole-like quasiparticles incoming 
from the left ($s = 1, 2$) or right ($s = 3, 4$) electrode.
In Fourier representation, where $E$ is the energy of the incoming quasiparticle
($|E|>\Delta$) and $E_n=E+nV$, we write
\begin{equation}\label{ppp}
\Psi_j^{(r)}(t) = \sum_n e^{- i E_n t} 
\left( \begin{array}{c} u_{jn}^{(r)} \\ v_{jn}^{(r)} \end{array} \right) 
\end{equation}
and introduce electron- and hole-type spinors,
\[
\Phi_{jn}^e \equiv \left( \begin{array}{c} u_{jn}^{(+)} \\ u_{jn}^{(-)} \end{array}
\right) , \quad \Phi_{jn}^h \equiv \left( 
\begin{array}{c} v_{jn}^{(+)} \\ v_{jn}^{(-)} \end{array} \right) .
\]
These spinors are defined on a chiral space, with the two entries
corresponding to the right- and left-moving parts.
{}From Eq.~(\ref{response}), we proceed to derive 
a first set of Andreev reflection 
matching equations for $\Phi_{jn}^{e,h}$,
\begin{eqnarray*}
\Phi_{Ln}^e &=& 
 \left( \begin{array}{cl} a_n & 0 \\0 & a_n^{-1} \end{array} \right)
\Phi_{Ln}^h + \delta_{n0} \, (u_E^2 - v_E^2) \,
\left( \begin{array}{c} \delta_{s1}/u_E \\ - \delta_{s2} /v_E \end{array}
\right) \\ \Phi_{Rn}^e &=&  
 \left( \begin{array}{cl} a^{-1}_n & 0 \\0 & a_n \end{array} \right)
\Phi_{Rn}^h + \delta_{n0} \, (u_E^2 - v_E^2) \,
\left( \begin{array}{c} - \delta_{s4}/v_E \\ \delta_{s3} /u_E \end{array}
\right) ,
\end{eqnarray*}
where $a_n = (E_n - \sqrt{(E_n + i 0)^2 - \Delta^2})/\Delta$.
Furthermore, the $u_E, v_E$ denote the  $n=0$ entries in Eq.~(\ref{ppp}).
Note that these matching equations are not modified by the presence of phonons,
see Ref.~\cite{final}.

Iterating Eq.~(\ref{eom2}) and averaging over the phonons, we find
\begin{equation} \label{eom2iter}
(i\partial_t - \Sigma)  \circ d = 
t_0\sigma_z \sum_{j=\pm,r} e^{\mp i \sigma_z Vt/2}\ \Psi_j^{(r)}(t) ,
\end{equation}
where $\circ$ is a shorthand for convolution. The phonon self energy
entering Eq.~(\ref{eom2iter})  is given by
\begin{equation}
\Sigma(t,t')=i\lambda^2 {\cal D}_0^{>} (t,t')\sigma_z {\cal G}_{0}^R(t,t')\sigma_z.
\end{equation}
Note that this ignores the polarization-bubble renormalization, see above.
For the four different scattering processes indexed by $s$, it is convenient to introduce 
matrices $\tilde{\Sigma}_{nm}$ according to
\begin{eqnarray}\nonumber
\tilde{\Sigma}(t,t') & = & 
e^{i \sigma_z \sigma_s t V/2} \, \sigma_z \, \Sigma(t,t') \,
\sigma_z \, e^{- i \sigma_z \sigma_s t' V/2} \\ \label{sss1}
& \equiv& \sum_{nm} \int_F \frac{d \omega}{2 \pi} \,  
e^{-i \omega_n t + i \omega_m t'} \, \tilde{\Sigma}_{nm}(\omega) 
\end{eqnarray}
with $\sigma_s = 1$ ($\sigma_s=-1$) for $s\in \{1,2\}$  ($s\in \{3,4\}$). 
To keep the notation simple, the $s$-dependence of $\tilde\Sigma_{nm}$
is not exhibited explicitly.

Using Eq.~(\ref{eom2iter}), we then obtain a second set of matching equations. 
For $s = \{1,2\}$,
\begin{eqnarray*}
\Phi^e_{R,n+1} &=& \left( T^e_n \right)^{-1} \, \Phi^e_{Ln} + 
\sum_m \left( S^{ee}_{nm} \, \Phi^e_{Lm} + 
S^{eh}_{nm} \, \Phi^h_{Lm} \right) \\
\Phi^h_{R,n-1} &=& \left( T^h_n \right)^{-1} \, \Phi^h_{Ln} + 
\sum_m \left( S^{hh}_{nm} \, \Phi^h_{Lm} + 
S^{he}_{nm} \, \Phi^e_{Lm} \right), 
\end{eqnarray*}
while for $s = \{3,4\}$,
\begin{eqnarray*}
\Phi^e_{L,n-1} &=& T^e_{n-1} \, \Phi^e_{Rn} + 
\sum_m \left( S^{ee}_{nm} \, \Phi^e_{Rm} + 
S^{eh}_{nm} \, \Phi^h_{Rm} \right)
\\ \Phi^h_{L,n+1} &=& T^h_{n+1} \, \Phi^h_{Rn} + 
\sum_m \left( S^{hh}_{nm} \, \Phi^h_{Rm} + 
S^{he}_{nm} \, \Phi^e_{Rm} \right) .
\end{eqnarray*}
Here, electron/hole transfer matrices are given by
\begin{equation}\label{transfer}
T_n^{e/h} = \left( \begin{array}{cc} 
1/t_n^{e/h} & \left( r^{e/h}_n / t^{e/h}_n \right)^\ast \\ 
r^{e/h}_n / t^{e/h}_n & 1/t_n^{e/h \ast} 
\end{array} \right) 
\end{equation}
with 
\[
t_n^{e/h} = - 1/(1 \mp i E_n^{e/h}/2\Gamma),
\quad r_n^{e/h}/t_n^{e/h} = \pm i E_n^{e/h}/2 \Gamma, 
\]
where $E_n^{e/h} = E+ (n \pm 1/2) V$.
Furthermore, transfer matrices linked to phonon-induced transitions are given 
in terms of the phonon self-energy matrix elements introduced in Eq.~(\ref{sss1})
\begin{eqnarray}\label{ttt}
\left( \begin{array}{cc} S^{ee} & S^{eh} \\ S^{he} & S^{hh} \end{array} \right)_{nm} &=& 
\frac{i\sigma_s}{2 \Gamma} ( \tau_z - i \tau_y )   \\ &\times& \nonumber
\left( \begin{array}{cc} 
- \tilde{\Sigma}^{11}(E) & \tilde{\Sigma}^{12}(E) \\ 
- \tilde{\Sigma}^{21}(E) & \tilde{\Sigma}^{22}(E) 
\end{array} \right)_{nm}  ,
\end{eqnarray}
where the Pauli matrices $\tau_{y,z}$ operate in the space 
of the $\Phi^{e,h}$ spinors and the $\alpha$ indices
 in $\tilde{\Sigma}^{\alpha\alpha'}$ refer to Nambu space.
Note again that these quantities all carry an implicit $s$-dependence.

Finally, after straightforward but somewhat tedious
algebra, the d.c.~current is expressed in terms of
the above scattering states,
\begin{eqnarray}\label{current}
I &=& e v_F \int_{|E| > \Delta} dE \frac{|E|}{\sqrt{E^2 - \Delta^2}} \  f_E 
\sum_{s = 1}^4  \sum_{n} \\ &\times& \nonumber
\sum_{a=e/h} \Bigl[ ( \delta_{s1} + \delta_{s2} ) 
 \Phi_{Rn}^{a \dagger}(E) \ \tau_z \Phi_{Rn}^a(E) \\ & +& \nonumber 
( \delta_{s3} + \delta_{s4} ) \Phi_{Ln}^{a \dagger}(E) \ \tau_z \Phi_{Ln}^a(E) \Bigr] .
\end{eqnarray}
Similarly, the full Keldysh Green function $\check {\cal G}$ can also be constructed from 
the complete set of scattering states $\Phi^{e,h}_{jn}(E)$.


\begin{thebibliography}{99}
\bibitem{foot}
The accuracy of the curve (see especially the left inset)
is somewhat compromised for large
$n$ because a voltage grid of constant stepsize has been used.
\end{thebibliography}
\end{document}